# Automated recognition of the pericardium contour on processed CT images using genetic algorithms

É.O. Rodrigues [a,*], L.O. Rodrigues [b], L.S.N. Oliveira [c], A. Conci [a], P. Liatsis [d]

[a] *Department of Computer Science, Universidade Federal Fluminense, Niterói, Rio de Janeiro, Brazil*
[b] *School of Pharmacy, Universidade Federal Fluminense, Niterói, Rio de Janeiro, Brazil*
[c] *School of Nursing, Universidade Federal Fluminense, Niterói, Rio de Janeiro, Brazil*
[d] *Department of Electrical and Computer Engineering, Khalifa University of Science and Technology, Petroleum Institute, PO Box 2533, Abu Dhabi, United Arab Emirates*

A B S T R A C T

This work proposes the use of Genetic Algorithms (GA) in tracing and recognizing the pericardium contour of the human heart using Computed Tomography (CT) images. We assume that each slice of the pericardium can be modelled by an ellipse, the parameters of which need to be optimally determined. An optimal ellipse would be one that closely follows the pericardium contour and, consequently, separates appropriately the epicardial and mediastinal fats of the human heart. Tracing and automatically identifying the pericardium contour aids in medical diagnosis. Usually, this process is done manually or not done at all due to the effort required. Besides, detecting the pericardium may improve previously proposed automated methodologies that separate the two types of fat associated to the human heart. Quantification of these fats provides important health risk marker information, as they are associated with the development of certain cardiovascular pathologies. Finally, we conclude that GA offers satisfiable solutions in a feasible amount of processing time.

## 1. Introduction

An increasing demand for medical diagnosis support systems has been observed jointly with increases in computing power in recent years. These systems speed up the tedious and meticulous manual analysis done by physicians or technicians on patients' medical data, where, in many cases, a huge amount of data requires processing and, therefore, the data supporting diagnosis may lack precision and suffer noticeable inter- and intra-observer variation.

Cardiac epicardial and mediastinal fats are correlated to several cardiovascular risk factors [1]. At present, three imaging modalities appear suitable for quantification of these adipose tissues, namely Magnetic Resonance Imaging (MRI), Echocardiography and Computed Tomography. Each of these modalities has been used in several works in the literature [2–4]. However, Computed Tomography provides a more accurate evaluation of fat tissues due to its higher spatial resolution compared to ultrasound and MRI. In addition, CT is widely used for computing the coronary calcium score.

In this work, we propose a simple yet robust method to automatically identify the pericardium contour of processed cardiac CT images. The pericardium appears as an elliptical object in the CT images of the axial-plane. Given images of the ground truth in Ref. [5], we are able to delineate the pericardium layer and separate the epicardial from the mediastinal fats. The proposed methodology is based on determining the parameters of an ellipse using Genetic or Evolutionary algorithms.

The methodology proposed in this work can (1) improve processing time for the automated segmentation of the cardiac fats. In a previous work [5], we proposed a method that automatically segments the epicardial and mediastinal fats on CT images using machine learning. However, this processing can take up to 1.8 h for a single patient, which corresponds to segmenting a total of 44 images on average. Instead, a single image can be processed by the previously proposed methodology. Next, the pericardium can be delineated and, thereafter, the traced ellipse can be propagated to the remaining 43 images, speeding up segmentation time considerably. The quantification of the epicardial fat can be performed by counting the amount of voxels inside the propagated ellipse. Furthermore, the method proposed in this work also (2) improves the visualization of the pericardium in the processed images of [5] and could also improve the accuracy of the obtained segmentation by disregarding incorrectly segmented epicardial fat.

* Corresponding author.
*E-mail addresses:* erickr@id.uff.br (É.O. Rodrigues), lucasor@id.uff.br (L.O. Rodrigues), larissasno@id.uff.br (L.S.N. Oliveira), aconci@ic.uff.br (A. Conci), pliatsis@pi.ac.ae (P. Liatsis).

## 2. Literature review

The pericardium is a fibroserous sac that contains the human heart. It is composed of three concentric layers: (1) the parietal layer, (2) the serous pericardium and (3) the fibrous pericardium, from the inner surface of the heart to the outermost layer, respectively. The pericardium separates two types of adipose tissues that are tightly associated to the human heart. The fat enclosed by the pericardium is usually called epicardial fat, whilst the outer fat is usually called mediastinal or pericardial fat [5].

A significant amount of studies correlate cardiovascular risk factors or conditions such as atherosclerosis [6–8], myocardial infarction [9], diastolic filling [10], atrial fibrillation and ablation outcome [11], carotid stiffness [12], etc [1,9,13–15], to the epicardial fat volume. Furthermore, the progression of coronary artery calcification is associated to the epicardial fat volume, as suggested by previous works [9,16]. Chen et al. [17] associate high coronary artery calcium score to a higher general cancer incidence.

Moreover, a 16-year study [18] that assessed a total of 384,597 patients found a rate of approximately 38.4% of deaths in the subsequent 28 days of individuals that had their first major coronary event. The same study also concludes that the occurrence of fatalities is slightly less associated to female individuals. Another study ranks cardiovascular incidents as the most common cause of sudden natural death [19]. Therefore, the practice of automatically evaluating the amount of fat related to the heart may contribute to avoid similar outcomes.

Automated quantitative analysis of the epicardial and mediastinal fats may provide a prognostic value to cardiac CT trials, delivering an improvement on its cost-effectiveness. Besides, that automation diminishes the variability introduced by different observers. In fact, quantifying these data by direct user interaction is highly prone to inter- and intra-observer variability. Thus, evaluated samples may not be associated to a unified sense of segmentation. Iacobellis et al. [20] have shown that epicardial fat thickness and coronary artery disease correlate independently of obesity. This evidence supports the independent segmentation and further quantification of adipose tissues rather than merely estimating their volume based on the overall fat of the patient.

In a previous work [5], we proposed an accurate method for automated segmentation of epicardial and mediastinal fats in cardiac CT exams. The technique is based on feature extraction [21], classification algorithms [22] and image dilation [23]. However, a necessary preprocessing step is image registration, which is performed prior to feature extraction. The downside of the approach is its processing time. Currently, approximately 1.8 h is required to automatically segment the cardiac fats of a patient, which consists of 44 images on average. However, no separation using ellipses is actually performed in that work.

We propose the automated identification and elliptical modelling of the pericardium layer, which separates the epicardial fat from the mediastinal fat. The identification is done using the ground truth provided in previous work [5], which can be found at [24]. Once the pericardium is identified in one of the images, it is possible to estimate the total volume of epicardial fat without segmenting every single image. This could dramatically reduce current total processing times from approximately 1.8 h to a few minutes [25]. We perform the identification by fitting an ellipse whose parameters are determined using a Genetic Algorithm [26].

Genetic or Evolutionary Algorithms are a group of computational optimizers inspired by Darwin's theory of evolution and its features such as natural selection, mutation, etc [27,28]. The main idea is to evolve a population of members that has intrinsic characteristics over time, where the good characteristics of individuals are preserved according to a fitness function. The characteristics of individuals could be, for instance, the parameters of a given problem. The solutions obtained with Genetic Algorithms may not be the best possible but are good enough, in general. The longer the period of evolution, the better the obtained results [29]. Metaheuristics such as Genetic Algorithms are much faster than brute force approaches, where all possible parameters must be evaluated.

GAs are stochastic search algorithms. This means that different solutions can be obtained if the seed for the pseudo-numbers generation is not preserved over different runs, even for the same problem. In its basic form, the starting point of a GA is a population of bit-strings, which are either called chromossomes or genotypes. Each bit-string represents a possible solution, called phenotype. The evolution of this population is performed over iterations or generations, at which chromossomes that inherit useful genetic traits are maintained in the population. As the generations proceed, several genetic operations can be used to simulate the evolution process, such as recombination of genes, asexual reproduction, serendipitous mutation, etc [27]. Eventually, the evolution process halts, according to the maximum desired number of generations or a predefined fitness threshold, and the best suited individual is selected as the result. Although the very first population is usually generated at random, they can be intelligently created if some characteristics of the solutions are known *a priori*. However, the evolution, based on the survival of the fittest, eventually leads to near-optimal solutions, even if the initial population is generated completely at random. The best fit solutions are granted more reproductive opportunities and, therefore, are qualified to disseminate their good characteristics to future generations. Other strategies, such as elitism, retain some of the best fit individuals to next generations [27].

Fig. 1 shows the generic steps of a GA. Starting from a population, individuals are selected based on their fitness function value. That is, individuals that have a greater fitness value will be selected more frequently. After the selection, genetic operations are applied or the individual is retained for the next generation. Eventually, a new individual is generated using genetic operators. In this case, its fitness function is computed and, based on this value, the algorithm decides whether the individual should be inserted in the population for future generations. This process is repeated until convergence is reached.

### 2.1. Related works

A number of semi-automated segmentation of epicardial fat have been proposed. Pednekar et al. [30] proposed a method for the segmentation of abdominal adipose tissues. The work of Bandekar et al. [31] further extended the method of Pednekar et al. to the segmentation of the epicardial fat. Coppini et al. [32] use a region growing strategy and a preprocessing step to remove all other thorax structures apart from the heart. An expert user is required to place control points along the pericardium border. Next, Catmull-Rom cubic spline functions are automatically generated to obtain a smooth pericardial contour. The volume of epicardial fat is obtained using thresholding. Although Coppini et al. [32] focus on reducing user intervention, the input of an expert is still required.

Barbosa et al. [33] further improved the automation of epicardial fat segmentation. They use the same preprocessing method proposed by Dey et al. [34] coupled with a high level step for identification of the pericardium. Identification is done by tracing lines originating from the heart's centroid to the pericardium layer and interpolating them using a spline curve. Although this approach is interesting, simple and highly applicable in virtually any of the previously described methods, the reported results demonstrate a low success rate. Only 4 out of 40 images were correctly segmented in a fully automatic way.

To the best of the authors' knowledge, Shahzad et al. [35] proposed the first fully automatic method for epicardial fat segmentation. Their method uses a multi-atlas based approach. This is based on registering several atlases (eight in this case) to a target patient and fusing these transformations to obtain the final result. No indication of the overall processing time is reported.

Ding et al. [36] proposed a similar approach. The pericardium is segmented using an atlas approach, which consists of minimizing the errors after applying transformations on the atlas along with an active contour method. The authors report that the atlases' images were pre-aligned to a standard orientation and hence, there is a comparison with only one of these atlases to speed up the process, which is a limitation.

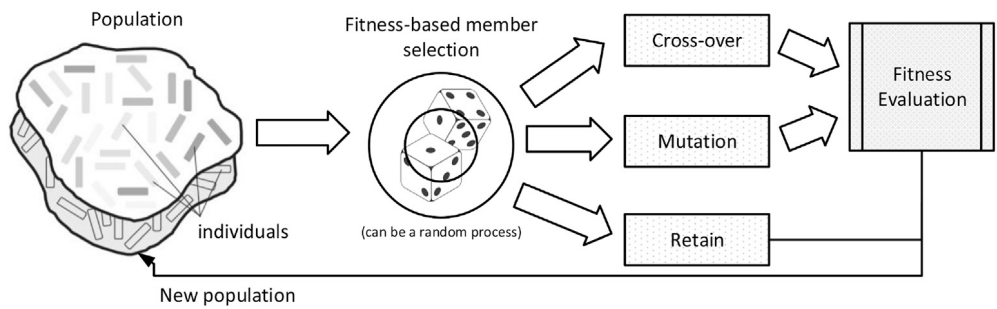

Fig. 1. Operation sequencing of a generic GA.

In a previous work [5], we proposed an automated methodology for segmentation of epicardial and mediastinal fats based on registration, classification and mathematical morphology. Although this method outperforms previously addressed works in terms of accuracy, it suffers from long processing times. It takes approximately 1.8 h to segment the CT volume information of a single patient.

The methodology proposed in this contribution provides for a means of locating the pericardium contour, hence enabling the estimation of the epicardial fat enclosed within the pericardium. Taking into consideration the smoothness constraint of the pericardium, the positional estimates of the pericardium contour on a single slice can be extrapolated to neighbouring slices, thus significantly improving processing times. Furthermore, the proposed pericardium tracing also reduces errors produced by segmentation methods, such as in Ref. [5], by reducing false positive epicardial fat voxels that may have been classified outside of the pericardium contour and also by reducing false negative voxels that may have been misclassified within the pericardium.

## 3. Materials and methods

The input images used in this work are Computed Tomography fat ranged images. The acquisition and representation of these images is explained in Ref. [5]. In theory, the images show just the fat tissues of the human body, i.e., the information within the $-200$ to $-30$ Hounsfield scale range. All other structures that are not related to fat are represented as black pixels, as shown in Fig. 2. The flowchart of the proposed methodology is shown in Fig. 3.

The ground truth [5] contains images of 20 patients, a total of 878 images, with the epicardial and mediastinal fats manually segmented. A single manually segmented image is shown in Fig. 4, where pixels labelled as red represent epicardial fat, green pixels represent mediastinal fat, grey pixels represent the remaining fats and black pixels represent the background.

Given images such as the one shown in Fig. 4, we aim to trace an elliptical contour matching with the pericardium, which is represented by the boundary region between the epicardial and mediastinal fats. The equation of the ellipse with rotation and displacement parameters is given by:

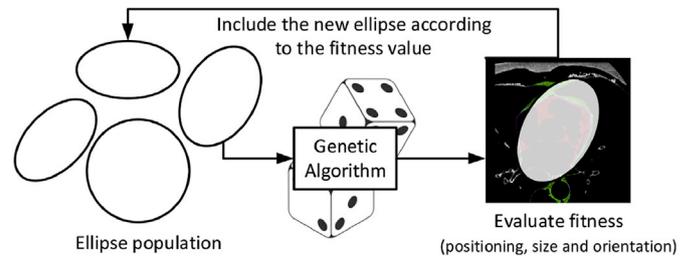

Fig. 3. An overview of the proposed methodology.

$$E(\theta, x_c, y_c, a, b) = \frac{((x-x_c)cos(\theta) + (y-y_c)sin(\theta))^2}{a^2} + \frac{((x-x_c)sin(\theta) - (y-y_c)cos(\theta))^2}{b^2} = 1 \quad (1)$$

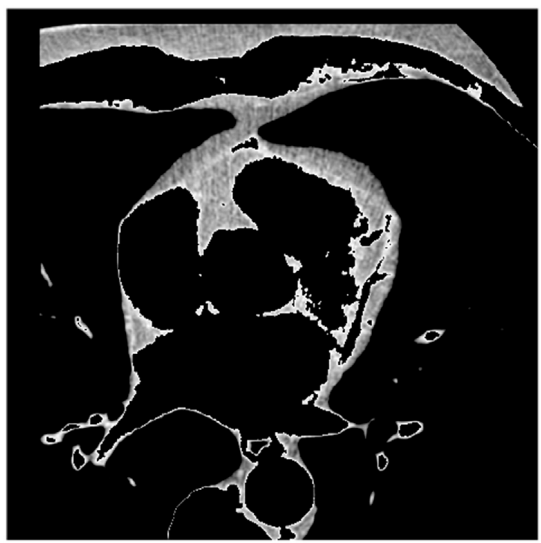

Fig. 2. A fat ranged Computed Tomography image on the axial plane.

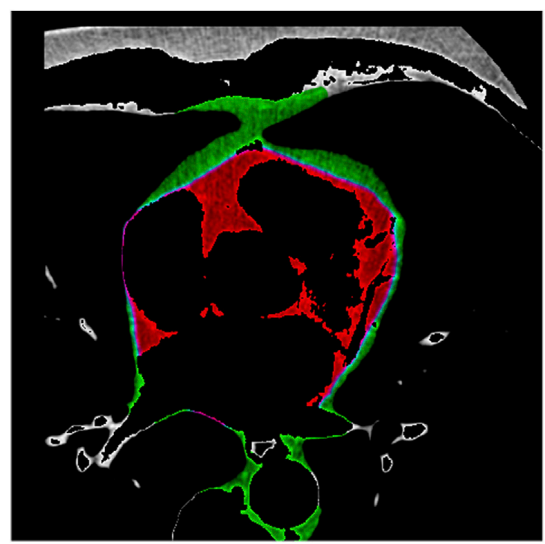

Fig. 4. A manually segmented image.

where $\theta$ represents the degree of rotation of the ellipse, $a$ and $b$ represent its major/minor axis, respectively, and $(x_c, y_c)$ stands for the ellipse center coordinates. The equation should be equal to 1 or, in the discrete case, equal to an interval such that $[1-\varepsilon, 1+\varepsilon]$, where $\varepsilon$ controls the size of the ellipse outline.

Our objective/fitness function is defined as $f_E$, shown in Equation (2). The objective function is a weighted score of red pixels $r$, green pixels $g$, grey pixels $c$ and black pixels $b$ within the ellipse $E$. The constants in the equation were empirically tuned using ground truth data, where $q_r = 85$, $q_g = 3$, $q_c = 4$, $q_b = 2.5$.

$$f_E = q_r r - (q_g g + q_c c + q_b b) \quad (2)$$

In detail, let the image $I$ contain pixels $p \in \mathbb{Z}^2$ and $R$, $G$, $C$ and $B$ be sets that contain the red, green, grey and black pixels of the processed image, respectively. The objective function is then given by:

$$f_E = \sum_{\forall p \in I} \begin{cases} \begin{cases} q_r, & \text{if } p \in R \\ -q_g, & \text{if } p \in G \\ -q_c, & \text{if } p \in C \\ -q_b, & \text{if } p \in B \end{cases} & \text{if } E(\theta, x_c, y_c, a, b) < 1 \\ 0, & \text{otherwise} \end{cases} \quad (3)$$

The red pixels are positively weighted since they are the ones to be maximized. In summary, the algorithm searches for ellipses $E$ that maximize the objective function $f_E$ within the image and uses a metaheuristic for this purpose.

The results obtained using Genetic Algorithms and other metaheuristics are based on random search, and thus will not necessarily provide the parameters of the optimum ellipse that yields the maximum of the objective function. However, as long as the Genetic Algorithm is well constructed and runs for a sufficient amount of time, the results would be satisfactory and sufficiently near to the optimal solution.

### 3.1. Genetic algorithm

The ellipse parameters are the orientation, center coordinates and major/minor axis as previously discussed. Thus, the individuals of the population contain five parameters encoded in their chromosome. Each parameter belongs to a predefined range, as given:

$$0 \leq \theta \leq 360, 100 \leq x_c \leq 412, 100 \leq y_c \leq 412, 80 \leq a \leq 300, 80 \leq b \leq 300,$$

where the input images have a size of $512 \times 512$ pixels. There is, therefore, an optimal set of parameters that produces an optimal ellipse for each image or slice of a single patient.

We maintain a population of 20 individuals in every generation. At each generation, the individuals that are best adapted, i.e., the ones that have the highest objective function values, remain in the population, while the remaining are discarded. At first, the 20 individuals are randomly generated. Starting at second generation, the children are the result of random crossovers between the individuals in the population and also mutations, as shown in Fig. 6.

We randomly select two individuals from the top 20. These two individuals generate a child. Each child inherits parameters from their parents, which are chosen randomly. At each generation or iteration, the top 20 individuals produce a total of 40 children, whose objective functions are evaluated to examine whether they should be inserted in the group of top 20 individuals. After crossover, random mutations may also occur, as previously mentioned. Algorithm 1 illustrates how crossover and mutation are performed, where the function $random(k)$ generates an integer number from 0 to $k-1$.

The crossover is merely a random pick with equal probability of parameters of the individual's parents. If parent A has $\theta = 360$ as parameter and parent B has $\theta = 56$, then there is an equal probability of the child having either one of them. This is respected for each parameter of the ellipse. After picking each parameter from one of the parents, the mutations may occur randomly, e.g., if a generated random number exceeds a given threshold.

**Data:** $\delta$ is the population (20 individuals), $n$ is the total amount of generations or iterations of the algorithm, $\Phi$ is a vector that contains the parameters of an individual $i$ such that $i.\Phi() = [\theta, x_c, y_c, a, b]$ (e.g., $i.\Phi(0) = \theta$, $i.\Phi(1) = x_c$, etc), $m$ controls the mutation and crossover rates, and $i_o$ represents a new individual

1  **method** geneticAlgorithm();
2  **begin**
3     Place 20 random individuals in $\delta$;
4     **while** *number of iterations* $< n$ **do**
5       **for** *a total of 40 children* **do**
6         Select two random individuals $i_1$ and $i_2$ from $\delta$;
7         Let $i_o$ be a clone of $i_1$;
8         **for** *k = 0, k<random(m), k++* **do**
9           r = random(5) ;   // 5 parameters
10          $i_o.\Phi(r) = i_2.\Phi(r)$;
11        **for** *k = 0, k<random(2 * m), k++* **do**
12          r = random(5) ;   // 5 parameters
13          $i_o.\Phi(r) += var()$;
14       Compute the objective function of $i_o$;
15       If the $i_o$ individual has an objective function that is higher than any individual in $\delta$, insert $i_o$ in the list and remove the worst one.
16    If $\delta$ has not changed in subsequent $\frac{n}{10}$ iterations, break;
17    Return the best individual in $\delta$;

**Algorithm 1.** Genetic algorithm for ellipse parameter selection.

Function $var()$, in Algorithm 1, is defined in Equation (4), where $randomf()$ generates a floating number between 0 and 1 (both inclusive). In summary, $var()$ essentially generates a random number between $-400$ and 400 if $u = 20$, where numbers near 0 are more often generated than numbers that are closer to $-400$ or 400. The constant $m$ is used to control the rate of mutation and cross over. This $m$ constant was empirically set to 5 and $u$ was set to 20.

$$var() = (random_i(u+1) \times randomf_i())^2 \\ - (random_j(u+1) \times randomf_j())^2 \quad (4)$$

Fig. 5 illustrates how the $var()$ function is computed. After generating a number $r_a$ using $random(u+1)$, where the possible values are illustrated by the two diagonal lines, $r_a$ is multiplied by a random floating number, such that the final value will be $\leq r_a$, as shown by the dashed vertical line. Finally, as a remark, the source code of the GA implementation is available at [37].

### 4. Results

The proposed methodology was applied to 3 axial plane images for each patient in the dataset. In total, an ellipse was generated for each of the 60 images. Besides, we carried out experiments by varying $n$, which corresponds to the maximum number of iterations or generations of the Genetic Algorithm. The indices used for analysis and comparison of the results were: (1) percentage of red pixels within the ellipse (PR), (2) percentage of green pixels within the ellipse (PG), (3) percentage of grey pixels within the ellipse (PC) and (4) percentage of black pixels within the ellipse (PB).

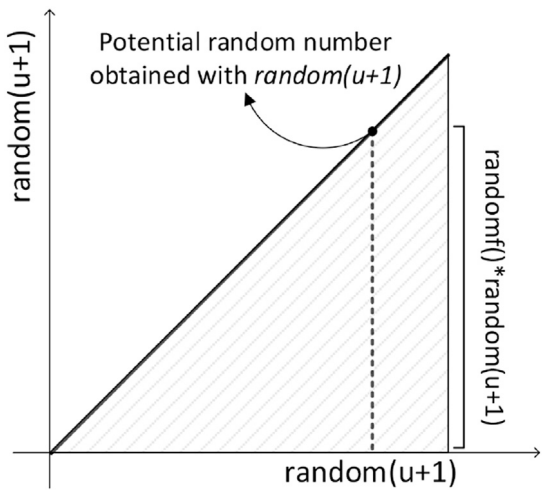

**Fig. 5.** Representation of the applied mutations using the *var* function.

We also developed an index that measures the fitness of the chosen ellipse as a whole by aggregating all of the previous indices (PR, PG, PC and PB). This index is called General Fit (GF) and defined as:

$$GF = \frac{PR}{PG + PC + PB} \qquad (5)$$

Table 1 shows the obtained indices for 10 generations of the Genetic Algorithm. In a real world scenario, we would perform pericardium detection on a single image of each patient, instead of 3 or even 50. The indices are presented as the mean, median, minimum and maximum values for all runs for the 20 patients. The last column corresponds to the processing time of the algorithm in seconds, i.e., for generating an ellipse in a single image using the GA. It is important to highlight that the algorithm was implemented in Java and was not optimized in regards to run times.

Table 2 presents the results for 100 generations, where the algorithm is run again from the start. The processing time increases by up to 10 times. However, the percentage of red pixels within the ellipse increases as well. The percentages of black, grey and green pixels decrease, which is a positive result. The percentage of green pixels, in particular, decreased substantially in comparison.

The red and green fats are located very close to each other and in some cases they are nearly impossible to separate, even for a human expert. However, it seems that in 10 generations, the algorithm captures the epicardial fat fairly well but aggregates a significant amount of mediastinal fat. The GF also improved substantially from 10 to 100 generations. We can fairly argue that the results improve in general if more iterations of the evolution are considered. However, there exists a tradeoff in this regard, this increases the processing time as well.

Finally, Table 3 shows the same indices for 200 generations. In this case, PR improved slightly when considering the results of Table 2 in relation to those of Table 1. However, the average processing time almost doubled. The averaged GF also improved in relation to 100 generations, but the improvement is not as significant as that from 10 to 100 generations.

Finally, some visual results for $n = 10$, 100 and 200 generations are shown in Figs. 7–9. The traced ellipse is represented in blue.

As previously mentioned, it is evident from these images that the results produced by 100 generations are already acceptable. Fig. 7 contains much more mediastinal and epicardial fat than Figs. 8 and 9. Besides, the area of the heart in this image is larger than the others. The images in Fig. 7 belong to CT image data obtained by a different CT scanner manufacturer, compared to the other two. Still, the approach managed to achieve good results in all occasions.

## 5. Discussion and future work

In a previous work [5], we proposed an automated process for segmentation of epicardial and mediastinal fats based on registration, classification algorithms and mathematical morphology. It extracts features from a pixel neighbourhood and uses this information to label the associated pixel. Classification algorithms are used as tools to decide whether a pixel belongs to the epicardial fat class. Following classification, dilations are performed to fill gaps in the segmentation performed by the classifiers.

Our previous research outperforms the two existing fully automated methods in the literature [35,36] in terms of accuracy. However, its main limitation is the processing time required to segment the patients' CT data. This registration, classification and mathematical morphology approach may take up to 1.8 h to process the data of a single patient. This is approximately 3 times faster than a human specialist, however further improvements are possible in order to achieve a near real-time

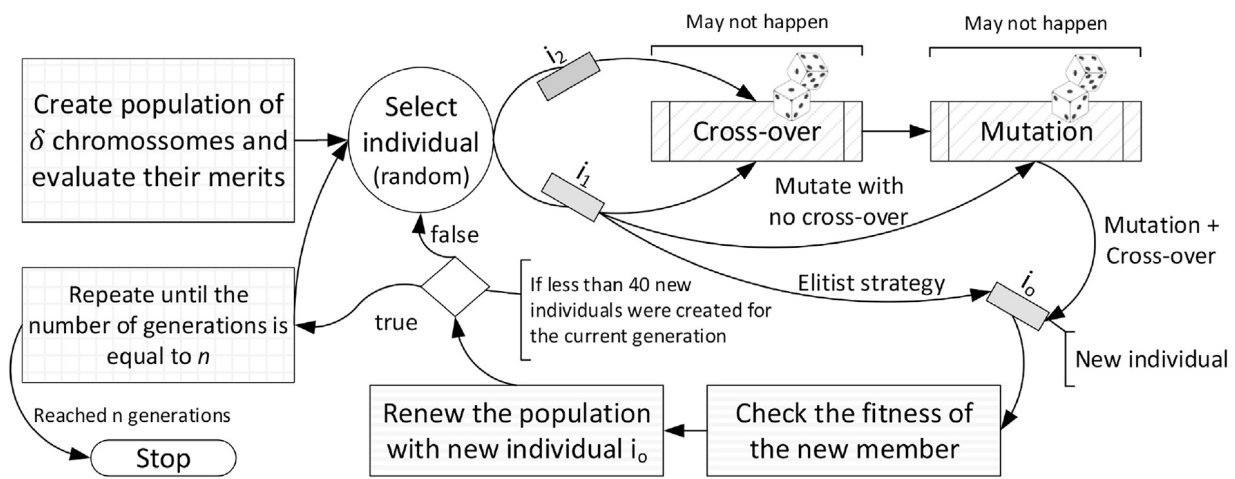

**Fig. 6.** Detailed steps of the proposed Genetic Algorithm.

**Table 1**
Results for 10 generations of the GA.

|        | PR    | PG    | PC    | PB    | GF   | Time (s) |
|--------|-------|-------|-------|-------|------|----------|
| Mean   | 97.31 | 44.86 | 6.00  | 20.47 | 1.95 | 38.43    |
| Median | 99.35 | 40.89 | 2.00  | 26.90 | 1.29 | 42.01    |
| Min    | 67.70 | 7.41  | 0     | 2.64  | 0.58 | 20.51    |
| Max    | 100   | 100   | 47.15 | 54.47 | 5.98 | 50.12    |

**Table 2**
Results for 100 generations of the GA.

|        | PR    | PG    | PC    | PB    | GF    | Time (s) |
|--------|-------|-------|-------|-------|-------|----------|
| Mean   | 98.85 | 25.60 | 1.64  | 16.43 | 3.16  | 378.72   |
| Median | 99.94 | 22.36 | 0.08  | 12.03 | 2.60  | 417.61   |
| Min    | 71.81 | 1.57  | 0     | 3.93  | 0.91  | 193.38   |
| Max    | 100   | 100   | 36.82 | 48.27 | 10.61 | 472.56   |

**Table 3**
Results for 200 generations of the GA.

|        | PR    | PG    | PC   | PB    | GF    | Time (s) |
|--------|-------|-------|------|-------|-------|----------|
| Mean   | 99.45 | 23.53 | 1.12 | 16.61 | 3.37  | 774.06   |
| Median | 99.95 | 17.59 | 0.12 | 11.68 | 3.02  | 819.99   |
| Min    | 92.73 | 1.63  | 0    | 3.51  | 0.9   | 334.93   |
| Max    | 100   | 100   | 9.3  | 47.69 | 10.54 | 935.08   |

performance.

In this context, an elliptical contour estimation procedure for the pericardium boundary is proposed in this work. This results in a performance improvement in our previously proposed framework [5] by, at first, improving accuracy by reducing false positive voxels outside of the pericardium contour and false negative epicardial fat voxels present within the pericardium contour. Furthermore, pericardium estimation can also speed up execution times.

Knowing the position and size of the pericardium ellipses on a few slices enables interpolation between them, thus providing an estimate of the epicardial fat within that axial section and improving time efficiency. That is, it will not be necessary to process every CT slice. For instance, a third of the CT slices could be processed instead, in which case the time required to process the cardiac CT examination of a single patient will reduce to approximately 0.6 h. In a previous work [38], we used regression algorithms to estimate the amount of mediastinal fat based on the epicardial and vice-versa. As such, the parameters of the pericardium ellipse may be propagated to neighbouring slices using regression algorithms [38], which could provide improved results compared to direct interpolation.

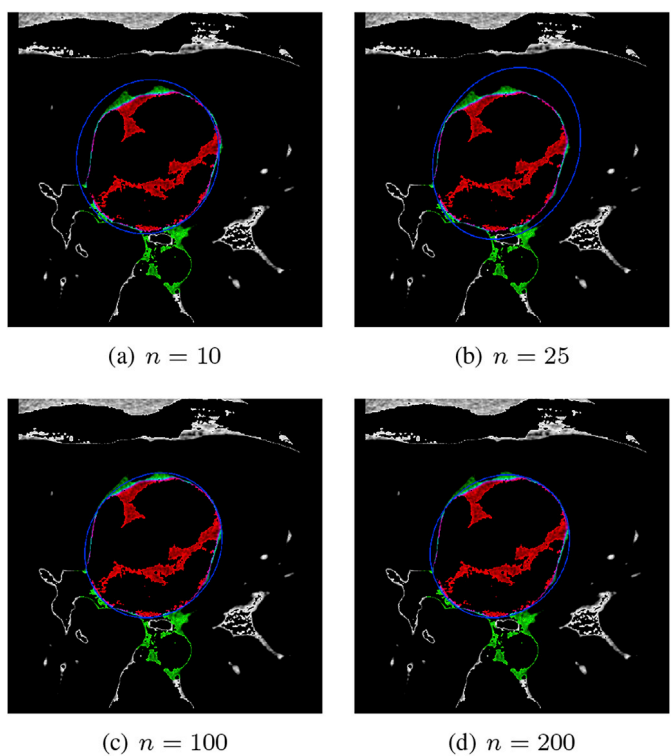

(a) $n = 10$   (b) $n = 25$

(c) $n = 100$  (d) $n = 200$

**Fig. 8.** Ellipses generated with GA for varying values of $n$.

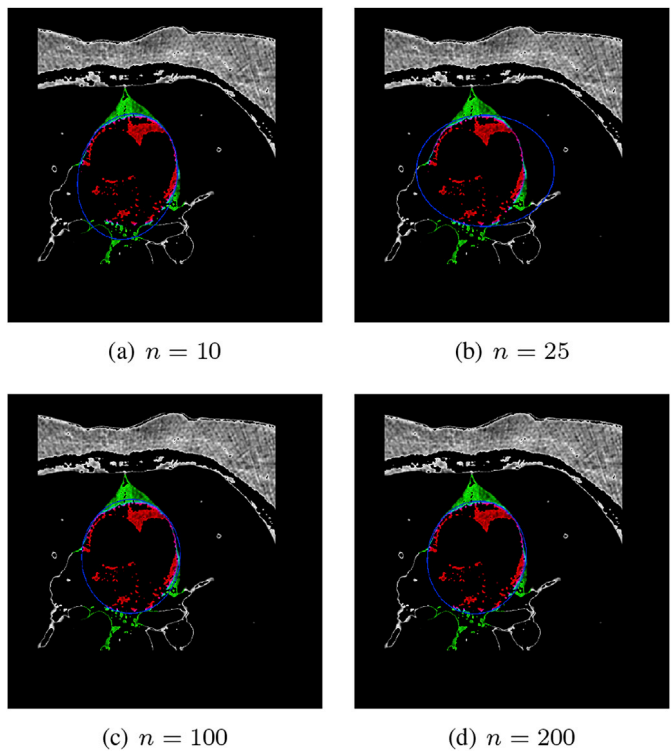

(a) $n = 10$   (b) $n = 25$

(c) $n = 100$  (d) $n = 200$

**Fig. 9.** Ellipses generated with GA for varying values of $n$.

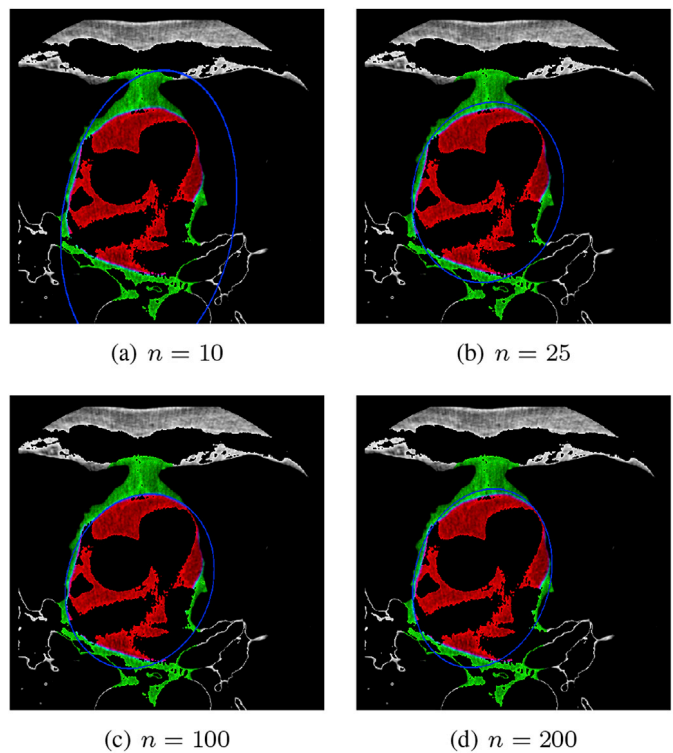

(a) $n = 10$   (b) $n = 25$

(c) $n = 100$  (d) $n = 200$

**Fig. 7.** Ellipses generated with GA for varying values of $n$.

In the case of regression algorithms, just two or three slices could suffice for the appropriate estimation of the fat volume regarding the entire slice spectrum. As observed in Ref. [38], it is possible to predict fairly accurately the amount of mediastinal fat in the entire cardiac CT examination, given that the information about the epicardial fat is known, using regression algorithms. Regression algorithms can be trained on patient data whose pericardium contour is known for the entire CT exam. Features such as the index of the slice and the parameters of the ellipse, including position, size and orientation, can be used in the regression analysis. The amount of epicardial fat can be the value to be predicted or the class of the problem. Once the regression algorithm is trained, the pericardium contours of a limited number of unprocessed CT images from a single patient can be extracted using the proposed methodology. These values, along with the index of an unprocessed slice, may be used to predict the volume of epicardial fat present in the respective slice.

In summary, the total estimation of epicardial fat can be done using the geometric information of the ellipse on a middle slice and adapting this to the preceding and following slices and to the information within so as to estimate the total amount of epicardial fat. Moreover, elliptical tracing is a visual measurement and a delineator that may assist physicians in detecting the pericardium, which is fairly difficult in some cases. As future work, we aim to explore the avenues previously described in this section, which consists of developing a means of estimating the epicardial fat volume of CT slices, hence reducing the time burden of the previously proposed segmentation approach.

## 6. Conclusions

In this work, we proposed the use of Genetic Algorithms to optimally fit ellipses in pre-processed Computed Tomography images in order to mathematically delineate the pericardium contour and, consequently, to separate two fats associated with the human heart, namely epicardial and mediastinal fats. Manual segmentation of these two fats and the pericardium is time consuming, tedious and generates extra costs. It would take approximately half a day to manually segment these two fats for a single patient. Due to these facts, quantification is commonly disregarded in usual medical practice.

The obtained results show that GA can find good solutions, i.e., closely trace the pericardium contour in a relatively fair amount of time. The algorithm can be further optimized in several ways and the processing time can be substantially reduced. For the experiments with 10, 100 and 200 generations, we achieved an average of 97.31%, 98.82% and 99.45% of epicardial fat being engulfed by the ellipse and fairly high GFs of 1.95, 3.16 and 3.37, respectively. For 200 generations ($n = 200$), we have visually confirmed accurate ellipse tracing for all of the 60 evaluated images.

## References

[1] G.A. Rosito, J.M. Massaro, U. Hoffman, F.L. Ruberg, A.A. Mahabadi, R.S. Vasan, C.J. O'Donnel, C.S. Fox, Pericardial fat, visceral abdominal fat, cardiovascular disease risk factors, and cardiovascular calcification in a community-based sample: the framingham heart study, Circulation 117 (2008) 605–613.
[2] R. Sicari, A.M. Sironi, R. Petz, F. Frassi, D.M. Chubuchny, V. Positano, M. Lombardi, E. Picano, A. Gastaldelli, Pericardial rather than epicardial fat is a cardiometabolic risk marker: an mri vs echo study, J. Am. Soc. Echocardiogr. 24 (10) (2011) 1156–1162.
[3] G. Iacobellis, M.C. Ribaudo, F. Assael, E. Vecci, C. Tiberti, A. Zappaterreno, U. DiMario, F. Leonetti, Echocardiographic epicardial adipose tissue is related to anthropometric and clinical parameters of metabolic syndrome: a new indicator of cardiovascular risk, J. Clin. Endocrinol. Metab. 88 (11) (2003) 5163–5168.
[4] T.S. Polonsky, R.L. McClelland, N.W. Jorgensen, D.E. Bild, G.L. Burke, A.D. Guerci, P. Greenland, Coronary artery calcium score and risk classification for coronary heart disease prediction, JAMA 303 (16) (2010) 1610–1616.
[5] E. Rodrigues, F. Morais, N. Morais, L. Conci, L. Neto, A. Conci, A novel approach for the automated segmentation and volume quantification of cardiac fats on computed tomography, Comput. Methods Programs Biomed. 123 (1) (2016) 109–128.
[6] A. Yerramasua, D. Dey, S. Venuraju, D.V. Anand, S. Atwal, R. Corder, D.S. Berman, A. Lahiri, Increased volume of epicardial fat is an independent risk factor for accelerated progression of sub-clinical coronary atherosclerosis, Atherosclerosis 220 (1) (2012) 223–230.
[7] R. Djaberi, J.D. Schuijf, J.M. Werkhoven, G. Nucifora, J.W. Jukema, J.J. Bax, Relation of epicardial adipose tissue to coronary atherosclerosis, Am. J. Cardiol. 102 (12) (2018) 1602–1607.
[8] T. Choi, N. Ahmadi, S. Sourayanezhad, I. Zeb, M.J. Budoff, Relation of vascular stiffness with epicardial and pericardial adipose tissues, and coronary atherosclerosis, Atherosclerosis 229 (1) (2013) 118–123.
[9] A.A. Mahabadi, N. Lehmann, H. Kälsch, T. Robens, M. Bauer, I. Dykun, S.M.T. Budde, K.H. Jöckel, S.M.R. Erbel, Association of epicardial adipose tissue with progression of coronary artery calcification is more pronounced in the early phase of atherosclerosis: results from the heinz nixdorf recall study, JACC Cardiovasc Imaging 7 (9) (2014) 909–916.
[10] S. Dabbah, H. Komarov, A. Marmor, N. Assy, Epicardial Fat, rather than Pericardial Fat, Is Independently Associated with Diastolic Filling in Subjects without Apparent Heart Disease, 2014, pp. 1–6.
[11] M.F.C.L. Schlett, M.F. Kriegel, F. Bamberg, B.B. Ghoshhajra, S.B. Joshi, J.T. Nagurney, C.S. Fox, Q.A. Truong, U. Hoffmann, Association of pericardial fat and coronary high-risk lesions as determined by cardiac ct, Atherosclerosis 222 (1) (2012) 129–134.
[12] T. Brinkley, F. Hsu, J. Carr, W. Hundley, D. Bluemke, J. Polak, J. Ding, Pericardial fat is associated with carotid stiffness in the multi-ethnic study of atherosclerosis, Nutr. Metab. Cardiovasc Dis. 21 (5) (2011) 332–338.
[13] R. Taguchi, J. Takasu, Y. Itani, R. Yamamoto, K. Yokoyama, S. Watanabe, Y. Masuda, Pericardial fat accumulation in men as a risk factor for coronary artery disease, Atherosclerosis 157 (1) (2001) 203–209.
[14] P. Raggi, Epicardial adipose tissue as a marker of coronary artery disease risk, J. Am. Coll. Cardiol. 61 (13) (2013) 1396–1397.
[15] P. Raggi, P. Alakija, Epicardial adipose tissue: a long-overlooked marker of risk of cardiovascular disease, Atherosclerosis 229 (1) (2013) 32–33.
[16] P.M. Gorter, A.M. de Vos, Y. Graaf, P.R. Stella, P.A. Doevendans, M.P.M.F.L. Meijs, F.L.J. Visseren, Relation of epicardial and pericoronary fat to coronary atherosclerosis and coronary artery calcium in patients undergoing coronary angiography, Am. J. Cardiol. 102 (4) (2008) 380–385.
[17] W. Chen, J. Huang, M.-H. Hsie, Y.-J. Chen, Extremely high coronary artery calcium score is associated with a high cancer incidence, Int. J. Cardiol. 155 (2012) 474–475.
[18] K. Dudas, G. Lappas, S. Stewart, A. Rosengren, Trends in out-of-hospital deaths due to coronary heart disease in Sweden (1991 to 2006), Circulation 123 (1) (2011) 46–52.
[19] C.T. Escoffery, S.E. Shirley, Causes of sudden natural death in Jamaica: a medicolegal (coroner's) autopsy study from the university hospital of the west indies, Forensic Sci. Int. 129 (2) (2002) 116–121.
[20] G. Iacobellis, E. Lonn, A. Lamy, N. Singh, A.M. Sharma, Epicardial Fat Thickness and Coronary Artery Disease Correlate Independently of Obesity, 2011, pp. 452–454.
[21] E. Rodrigues, A. Conci, T. Borchartt, A. Paiva, A.C. Silva, T. MacHenry, Comparing Results of Thermographic Images Based Diagnosis for Breast Diseases, 2014, pp. 39–42.
[22] E. Rodrigues, A. Conci, F. Morais, M. Perez, Towards the Automated Segmentation of Epicardial and Mediastinal Fats: a Multi-manufacturer Approach Using Intersubject Registration and Random Forest, 2015, pp. 1779–1785.
[23] E. Rodrigues, L. Torok, P. Liatsis, J. Viterbo, A. Conci, k-ms: a novel clustering algorithm based on morphological reconstruction, Pattern Recognit. 66 (2017) 392–403.
[24] V. Lab, A Ground Truth of Cardiac Fats, 2015. http://visual.ic.uff.br/en/cardio/ctfat/index.php.
[25] E. Rodrigues, V. Pinheiro, P. Liatsis, A. Conci, Machine learning in the prediction of cardiac epicardial and mediastinal fat volumes, Comput. Biol. Med. (2017), http://dx.doi.org/10.1016/j.compbiomed.2017.02.010.
[26] H. Santos, L. Ochi, E. Marinho, L. Drummond, Combining an evolutionary algorithm with data mining to solve a single-vehicle routing problem, Neurocomputing 70 (1) (2006) 70–77.
[27] J. Goulermas, P. Liatsis, Genetically fine-tuning the hough transform feature space, for the detection of circular objects, Image Vis. Comput. 16 (9–10) (1998) 615–625.
[28] P. Liatsis, Y. Goulermas, Minimal optimal topologies for invariant higher-order neural architectures using genetic algorithms, in: Proceedings of the IEEE International Symposium on Industrial Electronics, 1995.
[29] D. Whitley, A genetic algorithm tutorial, Stat. Comput. 4 (2) (1994) 65–85.
[30] A. Pednekar, A.N. Bandekar, I. Kakadiaris, M. Naghavi, Automatic segmentation of abdominal fat from ct data, Appl. Comput. Vis. 1 (2005).
[31] A. Bandekar, M. Naghavi, I. Kakadiaris, Automated pericardial fat quantification in ct data, Eng. Med. Biol. 1 (2006) 932–937.
[32] G. Coppini, R. Favilla, P. Marraccini, D. Moroni, G. Pieri, Quantification of epicardial fat by cardiac ct imaging, Open Med. Inf. 4 (2010) 126–135.
[33] J.G. Barbosa, B. Figueiredo, N. Bettencourt, J.M.R.S. Tavares, Towards automatic quantification of the epicardial fat in non-contrasted ct images, Comput. Methods Biomech. Biomed. Eng. 14 (10) (2011) 905–914.
[34] D. Dey, Y. Suzuki, S. Suzuki, M. Ohba, P.J. Slomka, D. Polk, L.J. Shaw, D.S. Berman, Automated quantitation of pericardiac fat from noncontrast ct, Investig. Radiol. 43 (2) (2008) 145–153.

[35] R. Shahzad, D. Bos, C. Metz, A. Rossi, H. Kirisli, A. van der Lugt, S. Klein, J. Witteman, P. de Feyter, W. Niessen, L. van Vliet, T. van Walsum, Automatic quantification of epicardial fat volume on non-enhanced cardiac ct scans using a multi-atlas segmentation approach, Med. Phys. 40 (2013) 9.
[36] X. Ding, D. Terzopoulos, M.D. Zamudio, D.S. Berman, P.J. Slomka, D. Dey, Automated epicardial fat volume quantification from non-contrast ct, in: Proceedings of SPIE, Medical Imaging 2014: Image Processing, 2014.
[37] E. Rodrigues, Source Code, 2017. https://github.com/Oyatsumi/PerdicardiumRecognition.
[38] E.O. Rodrigues, V.H. Pinheiro, P. Liatsis, A. Conci, Machine learning in the prediction of cardiac epicardial and mediastinal fat volumes, Comput. Biol. Med. (2017).